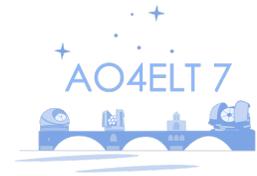

# High-Spectral Resolution Dark Holes: Concept, Results, and Promise

William Thompson[*a], Adam B. Johnson[b], Christian Marois[a], Olivier Lardière[a], Frédéric Grandmont[c], Tim Hardy[a], Kris Caputa[a], Colin Bradley[b], Garima Singh[a]

[a] NRC Herzberg Astronomy and Astrophysics, Victoria B.C., Canada;
[b] Department of Mechanical Engineering, University of Victoria, Victoria B.C., Canada;
[c] ABB, Quebec Q.C., Canada;
[d] NOIRLab, Gemini North Observatory, Hilo H.I., United States

**ABSTRACT**

Next generation high contrast imaging instruments face a challenging trade off: they will be required to deliver data with high spectral resolution at a relatively fast cadence (minutes) and across a wide field of view (arcseconds). For instruments that employ focal plane wavefront sensing and therefore require super-Nyquist sampling, these requirements cannot simultaneously be met with a traditional lenslet integral field spectrograph (IFU). For the SPIDERS pathfinder instrument, we are demonstrating an imaging Fourier transform spectrograph (IFTS) that offers a different set of tradeoffs than a lenslet IFU, delivering up to R20,000 spectral resolution across a dark hole. We present preliminary results from the SPIDERS IFTS including a chromaticity analysis of its dark hole and demonstrate a spectral differential imaging (SDI) improvement of up to 40×, and a first application of spectro-coherent differential imaging, combining both coherent differential imaging (CDI) and SDI.

## 1. INTRODUCTION

One of the main strengths of direct imaging and related techniques is the ability to study planets spectroscopically---that is, to measure the brightness of a planet at many wavelengths.

Spectrographs can be described along four axes:

- their wavelength range, or bandwidth;
- their spectral resolution, R;
- their field of view;
- and their noise properties.

To date, spectrographs in dedicated direct imaging instruments have fallen into two families.

The first family consists of high resolution, fibre-fed or long-slit spectrographs. These cross-dispersing spectrographs provide very high spectral resolution and bandwidth from a single location in the sky. They are effective instruments for characterizing known planets, provided that their positions can be accurately predicted in advance to within a small fraction of a λ/D.

The second family consists of low spectral-resolution imaging spectrographs, also known as Integral Field Spectrographs (IFS). These instruments capture a spectrum from each point in a grid of locations in the sky. An IFS is integrated into most major direct imaging instruments including the Gemini Planet Imager[1], VLT-SPHERE[2] and

SCExAO/CHARIS[3]. Because these instruments measure spectra at many locations, the planet does not have to be located or even known in advance. This makes IFS useful for detection in addition to planet characterization.

An IFS can be used to detect planets more effectively than a simple imager by searching for the spectral signal of a planet and/or by using spectral differential imaging (SDI [4]). Unfortunately, the potential of an IFS to search for planets using their spectral signals is limited by their low resolution. With a spectral resolution as low as R 50, only large molecular absorption bands stand out.

As it turns out, the limitations of both families (field of view or spectral resolution) have a common cause. Both designs are, at their heart, dispersive spectrographs and therefore require at least one detector pixel per spatial- and spectral-resolution element. With these designs, we are therefore limited by the size of astronomical detectors available for purchase. At the time of writing, appropriate detectors are available with at most 10s of millions of pixels. Very large format imagers like those for the Vera C. Rubin Observatory use rafts of multiple detectors [5] which come with complications and at considerable cost.

We now come to the Self Coherent Camera (SCC [6]). In addition to acting as an active wavefront sensor, an SCC allows one to perform coherent differential imaging as a post-processing step (CDI [7]–[15]). CDI is very powerful compared to ADI and SDI as it works using data from a single instant and wavelength. ADI is limited by speckle evolution while the pupil rotates and SDI is limited by chromatic evolution of the wavefront.

The SCC and therefore CDI were originally envisaged as operating only in a narrow bandpass. The multi-reference SCC concept extends this to a slightly wider bandpass in order to allow more light to reach the detector [16]; however, it still only records light from the planet with using a single bandpass and makes tradeoffs with respect to the reference beam intensity[14].

It is clearly desirable to develop a system that combines CDI with spectroscopy[1]. This would allow both access to more light like in the multi-reference SCC concept and, uniquely, allow for improved spectroscopic sensitivity by removing residual starlight from each wavelength slice in post-processing.

It is also worth noting the CDI is complementary to both ADI and SDI. All three could in theory be freely combined on the same dataset to achieve even deeper and more robust contrasts.

Unfortunately, a traditional dispersive IFS cannot effectively support the SCC or CDI. The SCC embeds the phase of the speckle electric field in high spatial-frequency fringes. To resolve these fringes, the detector must therefore super-Nyquist sample the scene. This would place at least four times greater pressure on already pixel-starved dispersive integral field spectrographs.

Besides the challenge of combining CDI with spectroscopy, an orthogonal challenge with dispersive integral field spectrographs may be chromaticity. Current instruments that support SDI are limited by chromatic evolution of the wavefront between spectral channels[17]. Simulations of known aberrations the Gemini Planet Imager's optics [18] do not fully account for this chromaticity. By contrast, laboratory experiments testing the use of tunable filters for SDI show no such loss in performance[19]. Accordingly, a plausible culprit may be lenslet integral field spectrographs themselves.

Lenslet based integral field spectrographs (sometimes known as IFUs) produce a grid of micro-spectra which are spread diagonally across the detector. To extract a spectral cube (2D intensity vs a third spectral axis), one must use software to interpolate along these micro-spectra. At high contrasts, it is plausible that pixel response, inter-pixel cross talk, and imperfect interpolation along the micro-spectra could lead to a loss of correlation between spectral channels. It is possible that this uncorrelated noise in turn limits the ability of SDI to suppress speckles. It is plausible that an integral field spectrograph built around a simpler extraction procedure could result in improved SDI processing.

## 2. PROPOSED SOLUTION

We propose that an imaging Fourier Transform Spectrograph would solve the challenges of accessing higher spectral resolution, combining CDI with spectroscopy, and of the supposed IFS chromaticity.

---

[1] Deformable mirrors can only perfectly correct a single wavelength so there is less reason to perform active wavefront control with multiple wavelengths compared to post-processing.

An IFTS is an imaging spectrograph built using a Michelson Interferometer.

Unlike a dispersive IFS, an IFTS records an entire spectrum on each pixel of the detector. This fundamentally removes the field of view—spectral resolution trade off.

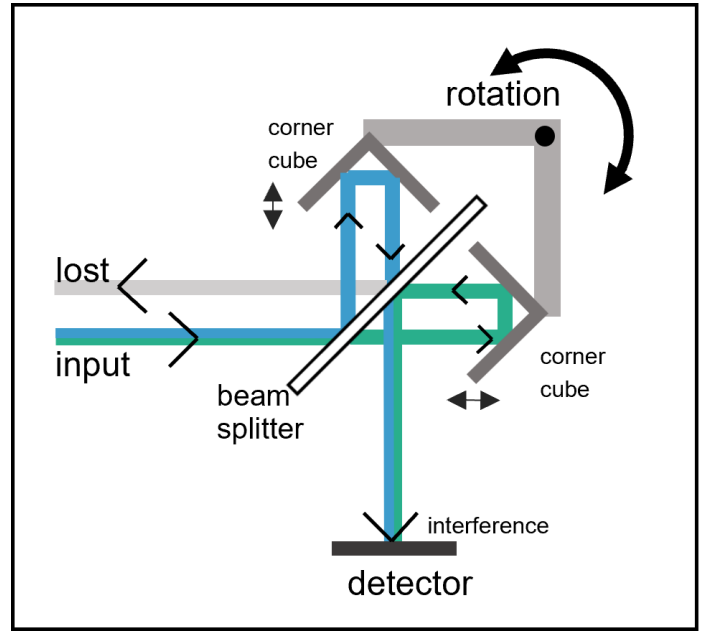

Figure 1. Schematic of an IFTS using a flexural bearing and corner-cube design [20].

In an IFTS, one directs a collimated beam of broad band light into an interferometer consisting of a 50/50 beam splitter and two mirrors with a varying optical path length difference. When one varies the length of one arm, light travels further down one arm than the other. When it then recombines through the beam splitter, it interferes with a delayed copy of itself. For a given optical path difference in nanometers, the phase difference in radians will depend on the wavelength of light. As such, scanning in optical path difference creates an intensity modulation that depends on the spectral content of the scene. This intensity vs. optical path difference (OPD) measurement is called an interferogram. One can recover the original spectrum simply by taking the discrete Fourier transform of the interferogram. Each pixel records its own interferogram and therefore its own spectrum.

An FTS has some similarities with a tunable filter (e.g. [19]), in that each detector pixel records an independent spectrum. It differs in that all wavelengths are observed simultaneously. This prevents any speckle evolution between wavelength slices, and results in a much higher observing efficiency. This difference also results in different noise behaviour. On the one hand, an FTS is much less sensitive to dark current and read noise than a tunable filter or dispersive spectrograph. Yet for the same reasons, it is more impacted by photon noise and photon-noise propagates between wavelength slices.

A schematic of an IFTS is presented in Figure 1. This design in based on a pair corner-cube retroreflectors. Unlike a traditional Michelson interferometer, both arms are fixed together, but pivot about a fixed beam-splitter. For a detailed review on corner cube IFTS, see [20].

The following subsection presents a simulated high resolution spectrum of an exoplanet (from Sonora models [21]), the resulting interferogram, and final recovered spectrum.

The benefits of an IFTS are significant. First, it is easy to accommodate the sampling requirements of CDI. Any detector capable of operating an SCC in a single-band should suffice. This would make it possible to measure the electric field of residual starlight speckles across multiple wavelengths. Second, it would allow for an IFS with 2-3 orders of magnitude higher spectral resolution. This would allow one to use cross-correlations with exoplanet atmosphere models to search for new planets and to detect particular compounds. Planetary RV should also be possible [22]. These are well beyond the capabilities of low-resolution IFSs used in dedicated direct imaging instruments.

IFTS have previously been deployed as astronomical instruments, though never behind an adaptive optics system or for high contrast applications. One example is the SITELLE spectrographs at the Canada France Hawaii Telescope [23]. SITELLE offers seeing limited, wide-field imaging-spectroscopy in the visible.

## 3. SIMULATIONS

In order to examine how an IFTS would perform in a direct imaging imaging instrument, we carried out a series of simulations. We used cloud-free Sonora Bobcat models [21] as a theoretical template spectrum. We selected a spectrum for a planet with solar metallicity and an effective temperature of 1500 K. We normalized the spectrum to have a J band relative contrast of $2\times10^{-4}$ around a $5^{th}$ magnitude star. This is somewhat optimistic as it would correspond to a very bright planet or a brown dwarf. We considered that the star had a flat spectrum that had been suppressed to a 1σ contrast of $1\times10^{-7}$. This is the raw contrast that is anticipated by a GPI CAL2-like system [14], [24], but would be optimistic for SPIDERS. We then considered that the planet was observed through an ideal optical filter allowing only light between 1.25 and 1.4 micron to pass. This template is shown in Figure 2.

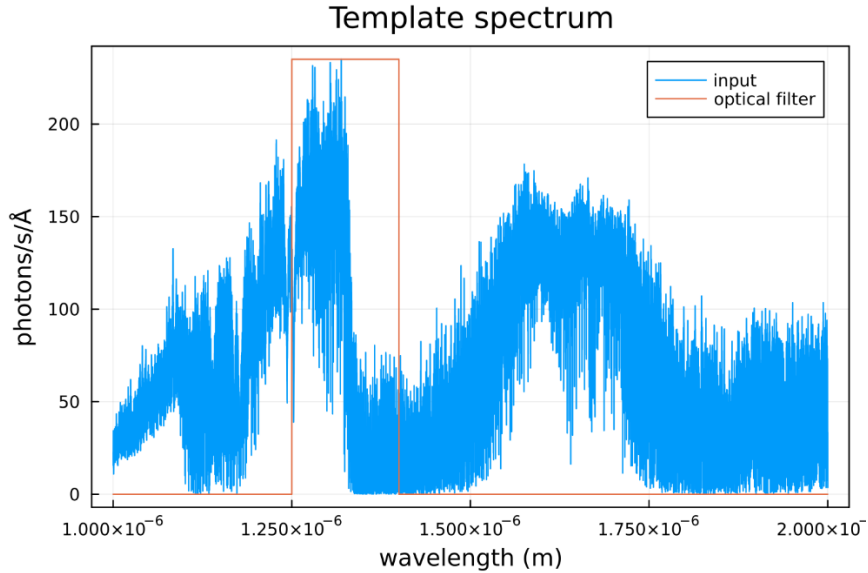

Figure 2. Template spectrum used for simulations. The overall intensity is normalized such that the planet has a J band relative contrast of $2\times10^{-4}$ around a $5^{th}$ magnitude star.

we further considered realistic illumination levels on an 8 m class telescope, 70% instrument throughput, and 65% modulation efficiency. We simulated a detector with very low read noise (0.5 $e^{-1}$/frame), 600 frames per second read out, and a scan step of 316 nm (half the wavelength of a Helium-Neon laser). This path length and step size results an average spectral resolution of R 30,000.

This scan speed was selected so that the intensity modulation we induce occurs at a higher temporal frequency than the majority of atmospheric turbulence. An alternative design would be to scan much more slowly so that variations in atmospheric speckles average during each exposure. Note that for space applications, a much wider range of scan speeds would be acceptable. The results of these simulations are presented in Figures 3-6.

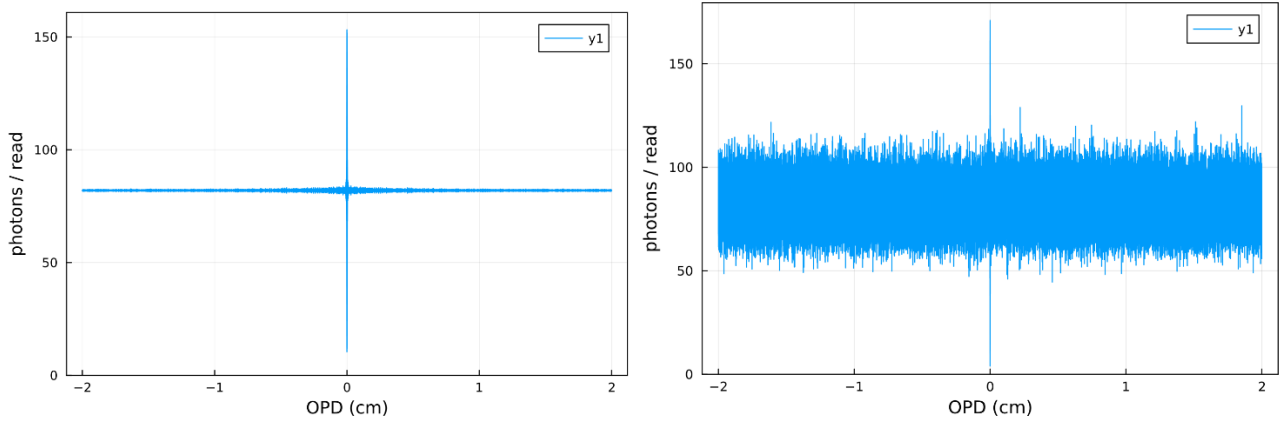

Figure 3. Simulated FTS interferogram for the parameters described in the text. The left panel shows an ideal, noiseless interferogram while the right panel shows an interferogram with photon and read noise.

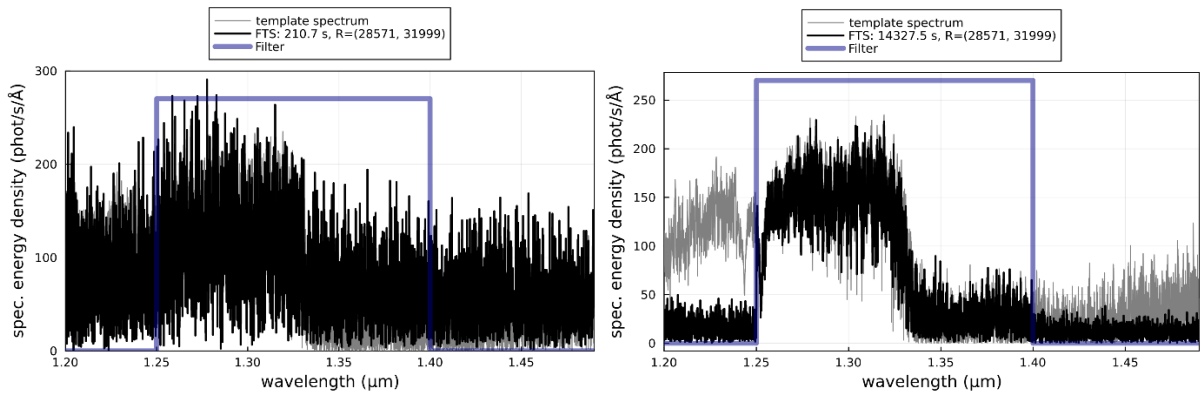

Figure 4. Simulated FTS spectrum for the parameters described in the text. The left panel shows a single ½ cm$^{-1}$ scan (R of approximately 30,000). For that resolution, a single scan takes 121 s. The right panel shows the same results stacked over four hours.

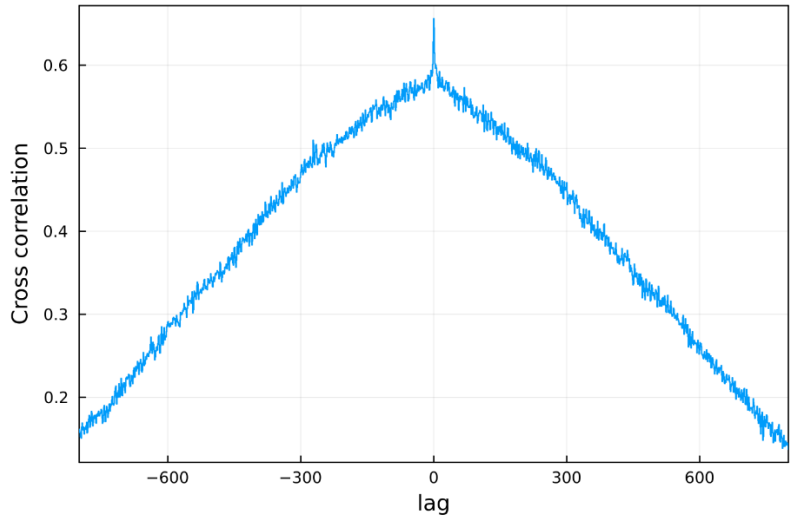

Figure 5. Cross-correlation of the template spectrum against a 1000 s integration of R 30,000 scans from the left panel of Figure 4. This demonstrates that the planet is still recovered robustly despite the noise. The SNR after performing a cross correlation depends on how accurately the model spectrum matches the planet's.

Beginning with the interferograms, Figure 3 shows an ideal case without any form of noise. The spike at zero-path length difference is the location where all wavelengths interfere constructively (called the ``centre-burst"). The narrow trough on either side of the centre-burst is the location where most of the light interferes destructively. If the template were a single narrow line, the interferogram would appear as a sine wave; however, since the template is relatively flat, most of the power is concentrated in a single peak in the interferogram. The second panel shows the same case including realistic (if slightly optimistic) noise levels. The simulations indicate that read noise has a negligible impact on the spectrum. Instead, quantization noise from the small number of Poisson distributed photons dominates.

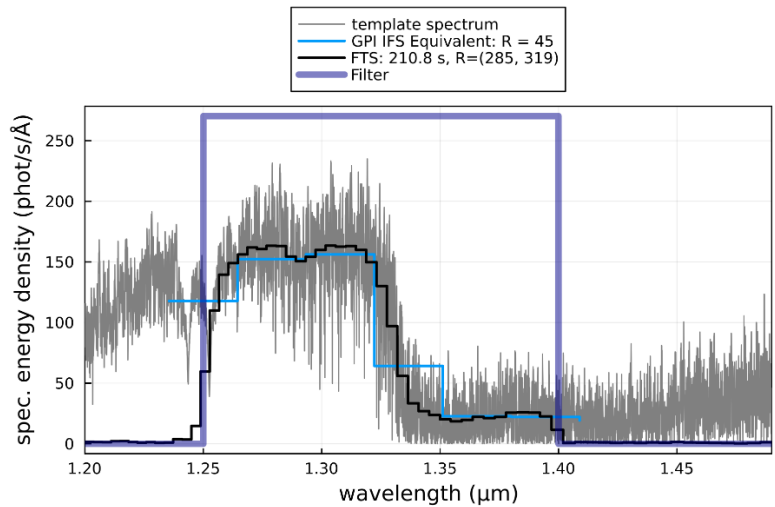

Figure 6. Simulations of low resolution FTS scans. As an alternative to the R 30,000 scan, this shows a stack of 100 scans captured R of approximately 300 for a total integration time of 3.5 min. This illustrates the flexible resolution of FTSs. The scan length should be chosen to maximize the amount of information retrieved.

Figure 4 shows spectra extracted from the simulation including realistic noise. These spectra have an average resolution of R 30,000. Due to the sheer faintness of planets, individual wavelength slices are noisy at this resolution. That said, the right panel illustrates a long integration of 4 hours. We see that the high resolution scan approaches the input spectrum. Despite the noise, Figure 5 shows that applying a cross-correlation between a 1000 s integration and the input template still results in a robust detection of the planet.

Finally, Figure 6 demonstrates a 3.5 minute integration of low-resolution scans (R of approximately 300). We see that the FTS recovers the broad features of the input template as expected. For the sake of comparison, a simulation of a low resolution dispersive spectrograph like that of the Gemini Planet Imager [1] is over-plotted.

The overall results of these simulations indicate that an IFTS would be a compelling alternative to a dispersive IFS for high contrast imaging.

## 4. IMPLEMENTATION

To build an imaging Fourier transform spectrograph, we repurposed a FTPA2000-300 series FT-NIR spectrometer. This line of spectrometers are used for recording the transmission spectra of gas samples. They contain a light source, Fourier Transform Spectrograph, and associated control electronics.

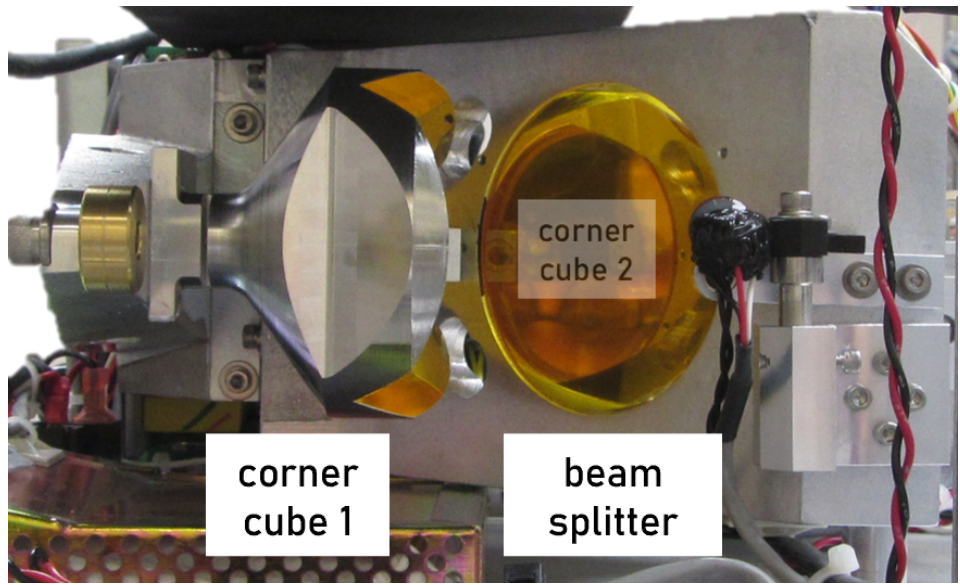

Figure 7. Image of the FTPA2000-300 series corner cube mirrors and beam splitter. In this perspective, the input beams enters into the page towards corner cube 2, and the output beam exits to the right.

We disassembled this device, keeping the FTS optics, voice-coil drive, and metrology laser system and discarding the light source, control electronics, and detector. Figure 7 shows an image of the disassembled FTS corner cubes and beam splitter. We built a new mechanical frame so that it could be fixed to an optical bench and optically aligned. Finally, we developed a new control electronics board that fully replaced the existing control system. This board interprets the signals of the remaining laser metrology system and drives a current through the devices voice coil actuator using a control loop. This custom control solution allowed us to scan the FTS at our own choice of scan speed, scan length, and trigger interval. These changes were essential since the original drive electronics scanned too quickly and would have required our camera detector to run at several thousand FPS.

For the tests presented in this chapter, we used a First Light Imaging CRED2 NIR detector. We configured our FTS controller to trigger the detector every 400 nm of optical path difference.

We used the resulting IFTS as a second science path in the SPIDERS pathfinder instrument [24]. This places the IFTS downstream of a coronagraph shared with a Self-Coherent Camera (SCC) focal plane wavefront sensor. A diagram of SPIDERS illustrating the SCC and IFTS paths is presented in Figure 8.

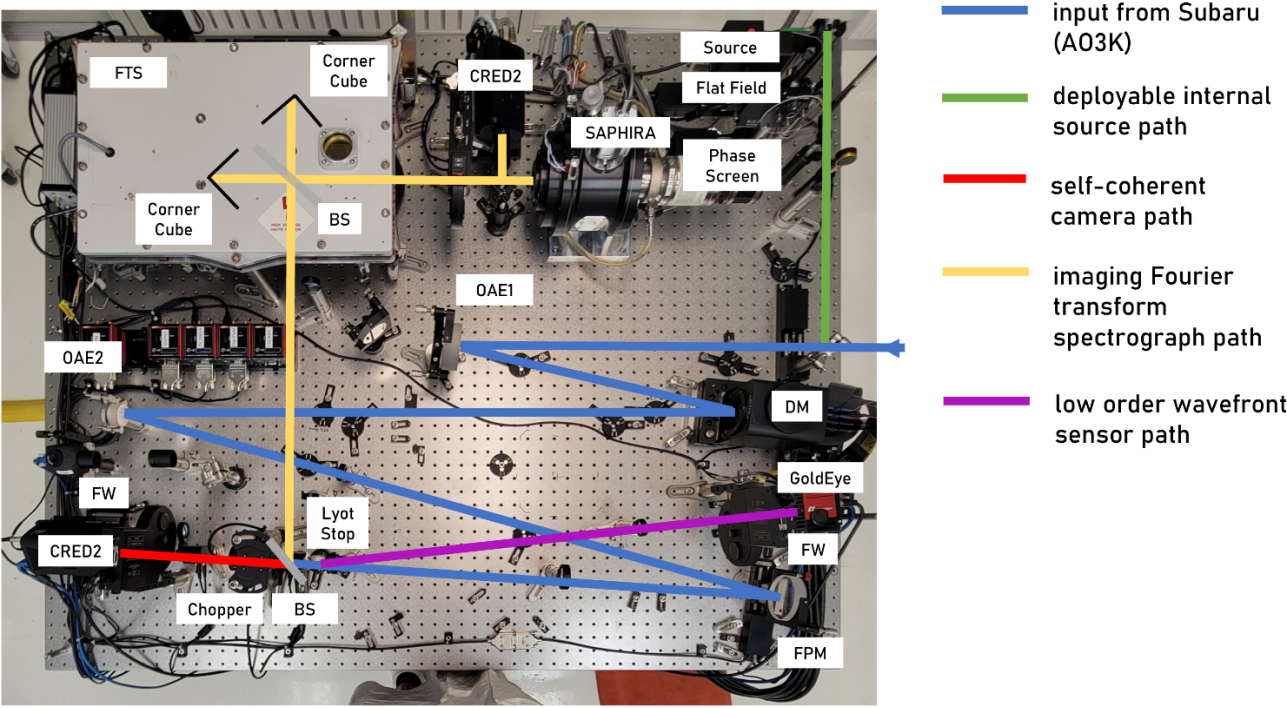

Figure 8. Diagram of the SPIDERS instrument showing the IFTS and SCC optical paths. The second CRED2 at the top of this image was used for these tests. The arrangement is similar to what is shown in Figure 7 with a corner cube at the top and at the left.

# 5. RESULTS

We now present preliminary results from the IFTS developed for SPIDERS. These experiments used an internal, pulsed white light source. The light source and camera (a First Light Imaging CRED2) were both triggered from the IFTS control electronics. A short section of the interferogram was used resulting in a spectral resolution of approximately R 350. This interferogram is presented in Figure 9.

Before capturing the IFTS data, we created a half dark hole using the SCC arm of the instrument with an H band apodized Lyot coronagraph and a narrow band filter. We paused the SCC control loop after the dark hole was created.

## 5.1 Interferograms and Spectra

Figure 10 shows image slices and a spectrum taken from a spectral data cube. The bottom panel shows the spectrum measured by summing over all pixels of each wavelength slice. A significant ripple in intensity is visible that is not yet explained at the time of writing, but may be a result of an optical ghost that travelled multiple times through the interferometer. Despite the unexpected intensity variations, the images are very clean and show the expected magnification with wavelength.

The first notable result is that the dark hole is relatively achromatic across this bandpass. This is highly significant as it means that narrowband correction by the SCC improves the surrounding wavelengths as well.

A second notable result is that the SCC fringes are sharply resolved in each wavelength slice as required.

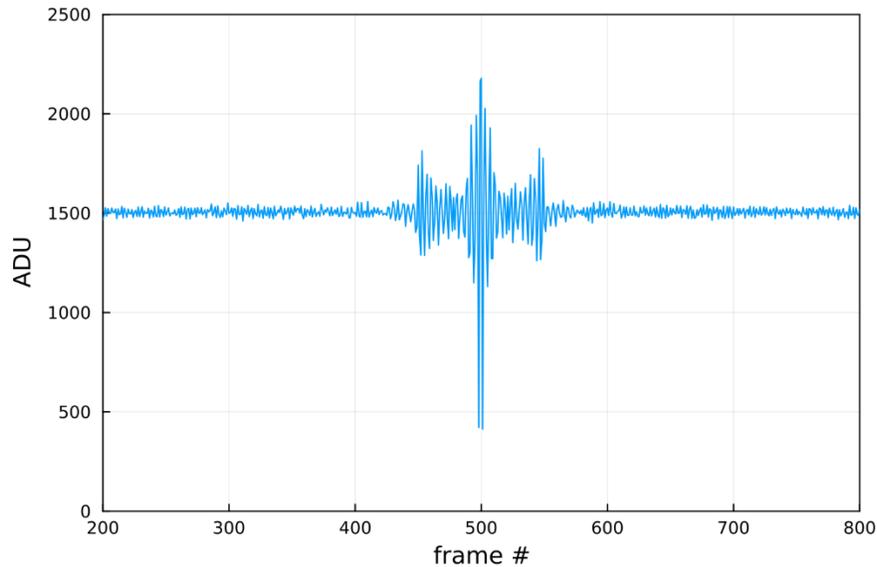

Figure 9. Interferogram recorded from the IFTS tests. For this plot, the data is summed over many camera pixels.

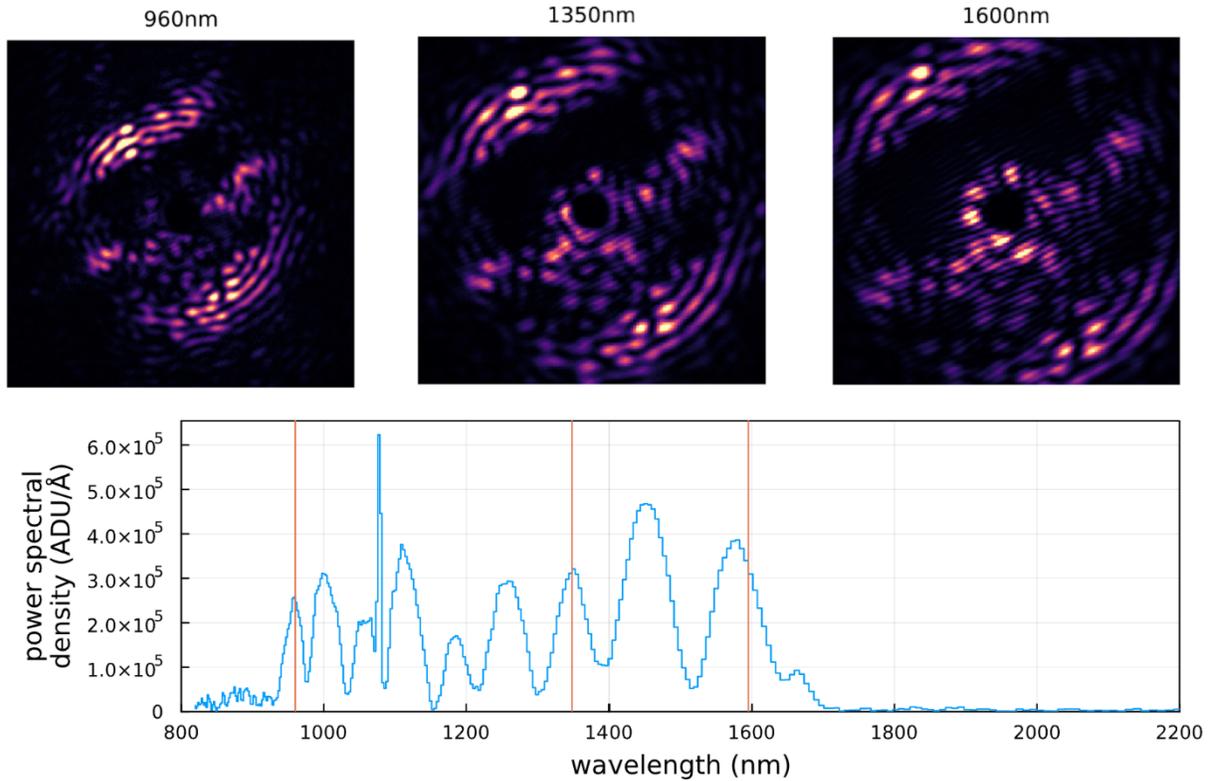

Figure 10. IFTS images from three different wavelength slices marked in the spectrum below. The cube is captured simultaneously across I, Z, Y, J, and H band. The input spectrum presented includes a large ripple feature due to a mirror with a dielectric coating optimized for visible reflectivity.

## 5.2 Spectral and Coherent Differential Imaging

Using these results, we attempted to perform SDI and CDI across all wavelength slices. For all cases, we optimized the subtraction in a rectangular region above the star that included the half dark hole and the surrounding speckles.

Figure 11 presents preliminary SDI results. For this experiment, we applied a simple global LOCI-style [25] algorithm (no sub-regions or annuli) that rejected slices taken 10% or closer in wavelength. We find that the extracted images are relatively achromatic over the wide 800 nm to 1650 mn bandpass. The contrast improvement, plotted in Figure 12, indicates that residual speckles outside the half dark hole are suppressed by up to nearly 40 times. This SDI result is highly promising as it exceeds the SDI improvement seen in GPI.

Thanks to the super-Nyquist sampling available in this design, the SCC fringes are resolved in each wavelength slice. This allows for the use of both SDI and CDI. Figure 13 then presents the application of both SDI and CDI on a given wavelength slice. For this case, we used the same least squares algorithm to the global SDI described above but added the CDI reconstructed image as an additional reference.

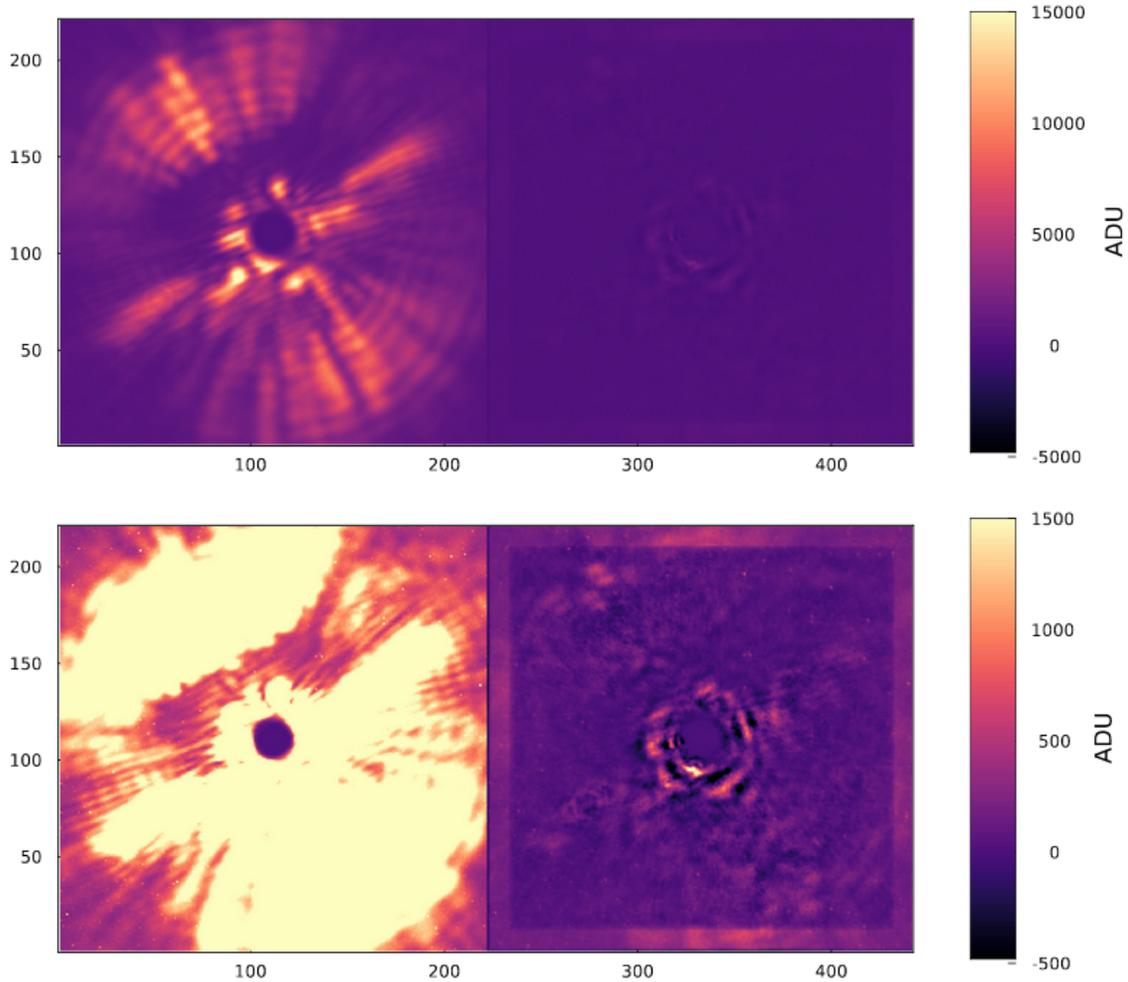

Figure 11. Results of applying SDI post processing to the IFTS data cube. The left panels show a broad band image stacked across all wavelengths shown in Figure 10 (IZYJH bands), while the right panels show the residuals after applying SDI. The bottom row presents the same data with intensity stretched by a factor of ten.

Both SDI and CDI perform very well on their own, and see a modest further improvement by combining the two. The level of this improvement approaches the noise floor of the images. In a real world situation, it is likely the CDI will have a greater impact than SDI at close separations. This is because the performance of SDI drops with decreasing separation, while the performance of CDI does not.

More experiments with realistic illumination levels, integration times, and atmospheric residuals will be needed to confirm that these techniques will work in a real world situation.

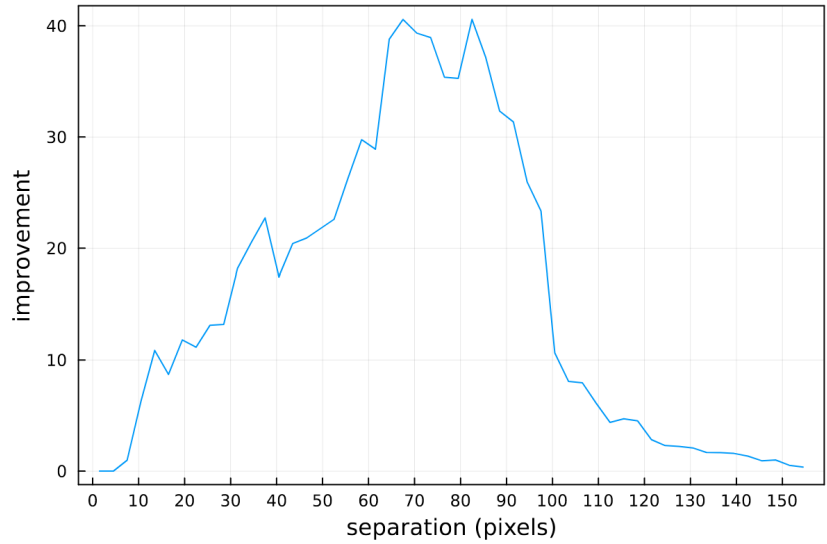

Figure 12. Contrast improvement from SDI applied to an IFTS data cube using a global 10% wavelength threshold.

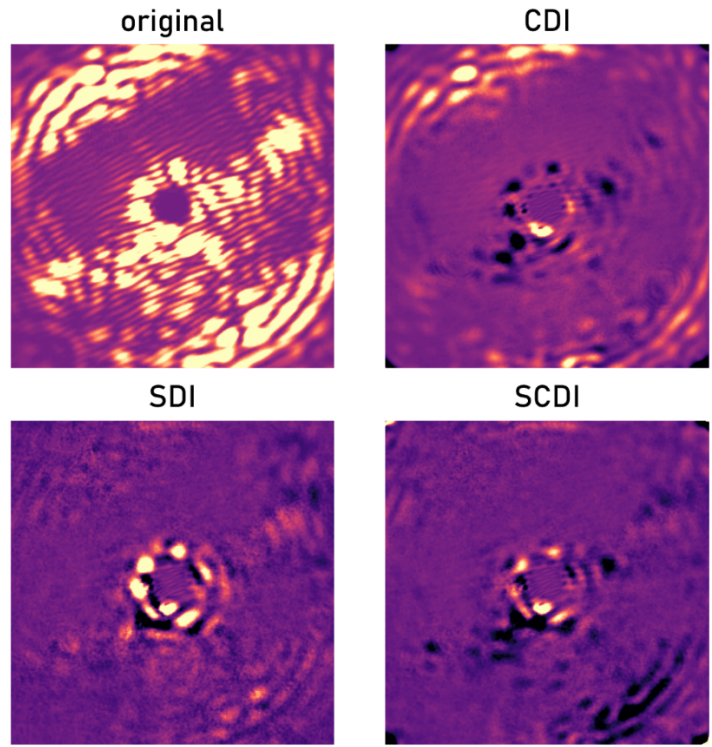

Figure 13. For a single wavelength slice (vs a stack in the previous figure), a comparison of CDI alone, SDI alone, and CDI and SDI working together. Both CDI and SDI offer a significant improvement in contrast and a modest further improvement is seen when combining them.

## 5.3 Dark Hole Darkness vs. Wavelength

Besides post-processing, these results also measure the chromaticity of the SPIDERS dark hole. Figure 14 presents compares the brightness of the speckles just outside the dark hole with the intensity of the residual light (without any post-processing) inside a half dark hole.

We find that the dark hole remains relatively dark across a factor of 2 in wavelength. This may be because our optical system is predominantly limited by phase errors instead of amplitude errors.

We used a coronagraph optimized for approximately 1500nm and find, as expected, that the residual dark hole intensity is minimized near this location. These early and preliminary results are not yet calibrated in contrast; only relative intensity is shown.

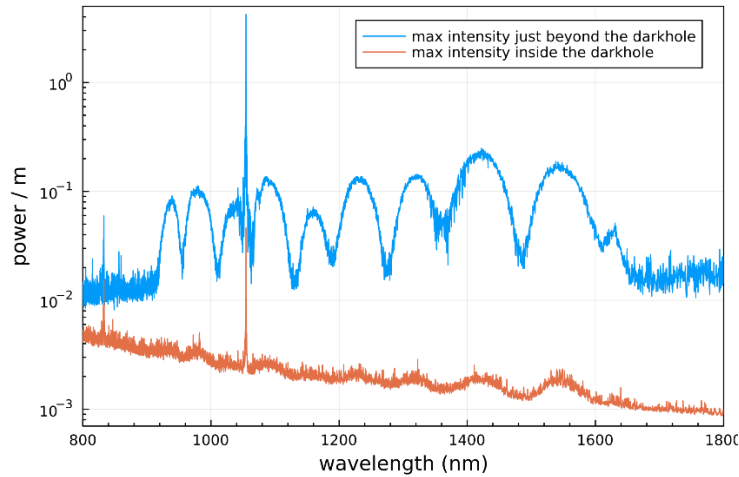

Figure 14. For a single wavelength slice (vs a stack in the previous figure), a comparison of CDI alone, SDI alone, and CDI and SDI working together. Both CDI and SDI offer a significant improvement in contrast and a modest further improvement is seen when combining them.

## 6. OUTLOOK

The simulation and lab results in this chapter show that an IFTS is a very promising spectrograph design for direct imaging. It can easily accommodate the sampling requirements of the SCC and CDI, while delivering much higher spectral resolution than a traditional dispersive IFU. That said, the very high spectral resolution provided by fibre-fed spectrographs continues to make them the best choice for detailed characterization of known planets.

This work demonstrates that a dark hole, constructed with focal plane wavefront sensing (the FAST SCC in this case), can be relatively achromatic, at least within the instrument itself. It also demonstrated that an IFTS can be used to improve contrasts with SDI beyond previous instruments, and with multi-spectral CDI. This later technique is particularly promising since its performance does not degrade at close separations from the star, nor does it negatively impact the signals of extended sources like debris disks.

Future instruments could incorporate an IFTS in the same path as the SCC to avoid splitting the light. This may be possible using a Wynne-corrector, an optical element that counteracts the magnification with wavelength effect. In such an instrument, each raw camera frame could be used to drive the SCC in broad band and real time (since all fringes would be aligned) all while the FTS is scanning. The resulting images could then be post processed to recover high resolution spectra.

# REFERENCES

[1] B. Macintosh *et al.*, "First light of the Gemini Planet Imager," *Proceedings of the National Academy of Science*, vol. 111, pp. 12661–12666, Sep. 2014, doi: 10.1073/pnas.1304215111.

[2] J.-L. Beuzit *et al.*, "SPHERE: the exoplanet imager for the Very Large Telescope," *Astronomy and Astrophysics*, vol. 631, p. A155, Nov. 2019, doi: 10.1051/0004-6361/201935251.

[3] M. W. McElwain *et al.*, "Scientific design of a high contrast integral field spectrograph for the Subaru Telescope," vol. 8446, p. 84469C, Sep. 2012, doi: 10.1117/12.927108.

[4] C. Marois, D. W. Phillion, and B. Macintosh, "Exoplanet detection with simultaneous spectral differential imaging: effects of out-of-pupil-plane optical aberrations," vol. 6269, p. 62693M, Jun. 2006, doi: 10.1117/12.672263.

[5] K. Gilmore *et al.*, "The LSST camera overview: design and performance," vol. 7014, p. 70140C, Jul. 2008, doi: 10.1117/12.789947.

[6] P. Baudoz, A. Boccaletti, J. Baudrand, and D. Rouan, "The Self-Coherent Camera: a new tool for planet detection," *Proceedings of the International Astronomical Union*, vol. 1, no. C200, pp. 553–558, Oct. 2005, doi: 10.1017/S174392130600994X.

[7] O. Guyon, "Imaging Faint Sources within a Speckle Halo with Synchronous Interferometric Speckle Subtraction," *The Astrophysical Journal*, vol. 615, pp. 562–572, Nov. 2004, doi: 10.1086/423980.

[8] P. Baudoz, A. Boccaletti, J. Baudrand, and D. Rouan, *The Self-Coherent Camera: a new tool for planet detection*. 2006, pp. 553–558. doi: 10.1017/S174392130600994X.

[9] P. J. Bordé and W. A. Traub, "High-Contrast Imaging from Space: Speckle Nulling in a Low-Aberration Regime," *The Astrophysical Journal*, vol. 638, pp. 488–498, Feb. 2006, doi: 10.1086/498669.

[10] A. Give'on, B. Kern, S. Shaklan, D. C. Moody, and L. Pueyo, "Broadband wavefront correction algorithm for high-contrast imaging systems," in *Astronomical Adaptive Optics Systems and Applications III*, SPIE, Sep. 2007, pp. 63–73. doi: 10.1117/12.733122.

[11] E. Serabyn, J. K. Wallace, and D. Mawet, "Speckle-phase measurement in a tandem-vortex coronagraph," *Applied Optics*, vol. 50, p. 5453, Oct. 2011, doi: 10.1364/AO.50.005453.

[12] J.-F. Sauvage, L. Mugnier, B. Paul, and R. Villecroze, "Coronagraphic phase diversity: a simple focal plane sensor for high-contrast imaging," *Optics Letters*, vol. 37, p. 4808, Dec. 2012, doi: 10.1364/OL.37.004808.

[13] B. L. Gerard, C. Marois, Raphaël. Galicher, and J.-P. Véran, "Fast focal plane wavefront sensing on ground-based telescopes," vol. 0703, p. 1070351, Jul. 2018, doi: 10.1117/12.2314134.

[14] B. L. Gerard, "Exoplanet imaging speckle subtraction: current limitations and a path forward," Thesis, 2020. Accessed: Nov. 16, 2020. [Online]. Available: https://dspace.library.uvic.ca//handle/1828/11755

[15] A. Potier *et al.*, "Increasing the raw contrast of VLT/SPHERE with the dark hole technique. II. On-sky wavefront correction and coherent differential imaging," *Astronomy and Astrophysics*, vol. 665, p. A136, Sep. 2022, doi: 10.1051/0004-6361/202244185.

[16] J.-R. Delorme, R. Galicher, P. Baudoz, G. Rousset, J. Mazoyer, and O. Dupuis, "Focal plane wavefront sensor achromatization : The multireference self-coherent camera," *A&A*, vol. 588, p. A136, Apr. 2016, doi: 10.1051/0004-6361/201527657.

[17] B. L. Gerard *et al.*, "A Chromaticity Analysis and PSF Subtraction Techniques for SCExAO/CHARIS Data," *The Astronomical Journal*, vol. 158, p. 36, Jul. 2019, doi: 10.3847/1538-3881/ab21d4.

[18] C. Marois, J.-P. Véran, and C. Correia, "A Fresnel propagation analysis of NFIRAOS/IRIS high-contrast exoplanet imaging capabilities," vol. 8447, p. 844726, Jul. 2012, doi: 10.1117/12.926826.

[19] P. Ingraham, "Détection et caractérisation de naines brunes et exoplanètes avec un filtre accordable pour applications dans l'espace," Mar. 2013, Accessed: May 23, 2023. [Online]. Available: https://papyrus.bib.umontreal.ca/xmlui/handle/1866/9194

[20] F. Grandmont, "Développement d'un spectromètre imageur à transformée de Fourier pour l'astronomieDéveloppement d'un spectromètre imageur à transformée de Fourier pour l'astronomieDevelopment of an imaging spectrometer with Fourier transform for astronomy;," 2006. Accessed: May 15, 2023. [Online]. Available: https://ui.adsabs.harvard.edu/abs/2006PhDT.......209G

[21] M. Marley *et al.*, "Sonora Bobcat: cloud-free, substellar atmosphere models, spectra, photometry, evolution, and chemistry." Zenodo, Jul. 14, 2021. doi: 10.5281/zenodo.5063476.

[22] J.-B. Ruffio *et al.*, "Detecting Exomoons from Radial Velocity Measurements of Self-luminous Planets: Application to Observations of HR 7672 B and Future Prospects," *AJ*, vol. 165, no. 3, p. 113, Mar. 2023, doi: 10.3847/1538-3881/acb34a.

[23] L. Drissen *et al.*, "SITELLE: An Imaging Fourier Transform Spectrometer for the Canada-France-Hawaii Telescope," *Monthly Notices of the Royal Astronomical Society*, vol. 485, no. 3, pp. 3930–3946, May 2019, doi: 10.1093/mnras/stz627.

[24] C. Marois *et al.*, "Deployment of focal plane WFS technologies on 8-m telescopes: from the Subaru SPIDERS pathfinder, to the facility-class GPI 2.0 CAL2 system," in *Adaptive Optics Systems VIII*, SPIE, Aug. 2022, pp. 594–603. doi: 10.1117/12.2630564.

[25] C. Marois, B. Macintosh, and J.-P. Véran, "Exoplanet imaging with LOCI processing: photometry and astrometry with the new SOSIE pipeline," in *Adaptive Optics Systems II*, International Society for Optics and Photonics, Jul. 2010, p. 77361J. doi: 10.1117/12.857225.